# Polar Order in Quantum Paraelectric SrTi$^{16}$O$_3$ and SrTi$^{18}$O$_3$ at Low Temperature

Yoshiaki Uesu[a),b)], Ryuhei Nakai[a)], Jean-Michel Kiat[c),d)], Carole Ménoret[c)], Mitsuru Itoh[e)] and Toru Kyomen[e)]

*a) Department of Physics, Waseda University, 3-4-1 Okubo, Shinjuku-ku, Tokyo 169-8555*
*b) Advanced Research Institute of Science and Engineering, Waseda University, 3-4-1 Okubo, Shinjuku-ku, Tokyo 169-8555*
*c) Laboratoire Structures, Propriétés et Modélisation des Solides, Ecole Centrale Paris, 92295 Châtenay-Malabry Cedex, France*
*d) Laboratoire Léon Brillouin, CE Saclay, 91191 Gif-sur-Yvette Cedex, France*
*e) Materials and Structures Laboratory, Tokyo Institute of Technology, 4259 Nagatsuda, Midori, Yokohama 226-8503*

Optical second-harmonic generation (SHG) in SrTi$^{16}$O$_3$ (STO16) and SrTi$^{18}$O$_3$ (STO18) was investigated using the SHG microscope. While no-biased STO16 exhibits weak and almost temperature-independent SHG signals, a marked SHG is observed under the electric field in the quantum paraelectric region. In STO18, strong SHG signals appear spontaneously below 36K. However, neutron and X-ray diffraction analyses indicate that no structural change appears at low temperature in STO18, and STO16 under the electric field. By taking into account the fact that the SHG is sensitive to the local polar-order, the combined studies reveals that the long-range order of polar phase does not develop on the both crystals and is frozen in local regions.

SrTiO$_3$ is a textbook example of both fields of fundamental and application of the solid state physic and new results and renewals of interests have continuously appeared. The most recent ones were initiated by the discovery by Itoh et al.[1] of a 'ferroelectricity' induced by the isotopic substitution of oxygen 18($^{18}$O) for oxygen 16($^{16}$O). Indeed, the progressive substitution of $^{18}$O in SrTi$^{16}$O$_3$ (STO16) which is a well-known quantum paraelectric with a quite high value of the dielectric constant (2x10$^4$) saturating at 0K [2], induces a broad maximum at a non zero temperature. Also, development of weak spontaneous polarization and D vs E hysteresis loops were observed below the peak temperature of dielectric constant [3]. This situation seems very similar to that encountered in STO16 under electric field above a critical field of 2kV/cm [4] and this phenomenon was also interpreted by the onset of ferroelectricity. However the existence of dielectric peak itself is not sufficient condition for the onset of ferroelectricity, as in the case of relaxor PbMg$_{1/3}$Nb$_{2/3}$O$_3$, where the development of polarization is restricted only in nano-scale regions whereas very strong value of permittivity is observed [5]. Furthermore, even near 0K, the dielectric constant of STO18 is quite high, approximately 2x10$^4$. This means that large dipole fluctuations still exist below the temperature of the maximum dielectric constant. These facts seem to suggest that the long-range order is not developed in STO18 even near 0K, as well as STO16 under the electric field. This is the motivation of the present research. For this purpose, we have performed optical second harmonic generation (SHG) experiments and neutron and high resolution X-ray diffraction experiments at low temperature in STO16 under electric field and in $^{18}$O-exchanged STO (STO18). This combination of studies allowed us to perform a comprehensive study of the polar order of these compounds at different length scales.

We used an SHG microscope [6,7] which makes images of SH waves produced in a polar specimen. In these experiments the magnitude and anisotropy of nonlinear optical tensor components $d_{ijk}$ was used in order to produce two-dimensional SH images of a specimen with inhomogeneous distribution of the polar region. The fundamental wave of Nd$^{3+}$:YAG laser with wavelength of 1063 nm, repetition frequency of 10 Hz and maximum fluence of 80mJ/cm$^2$, passes through a half-wave plate and is weakly focused into a specimen placed inside a helium gas exchange type cryostat for optical microscopes. The generated SH waves are collected by an objective lens and passes through a polarizer, filters and are detected by an image-intensified charge-coupled device (ICCD) camera [6]. In the present study, we used (110) plates with the edges along [1-10] and [001] directions. The dimensions of the plates were 5x5x0.25(thickness) mm$^3$ for STO16 and 7x3x0.3 mm$^3$ for STO18. The electric field was applied along the [1-10] direction. The exchange ratio of $^{18}$O in STO18 was 95%. The observation was made using the parallel polarizer-analyser system and the polarization direction of the incident wave was fixed with an angle making 45° from the [1-10] direction. As remarkable photoconductivity appears below 35K in STO16 [8], special cares were paid to avoid the effect: The diameter of the incident laser beam (2mm) was made smaller than the distances of two electrodes (4mm).

Some SH images of STO18 on zero-field cooling (ZFC) process are shown in Fig.1, where bright regions are SHG-active regions. A rather intense SHG appears below 36K. The images consist of small bright spots inhomogeneously distributed in the sample.

Judging from the size (several μm), the small bright spot does not directly correspond to a polar micro-region but aggregated ones. However the most important fact is that SHG inactive regions exist as dark regions. As SHG images are quite homogeneous in a domain of typical ferroelectrics, the images strongly suggest that the long range order does not develop in STO18 without the electric field.

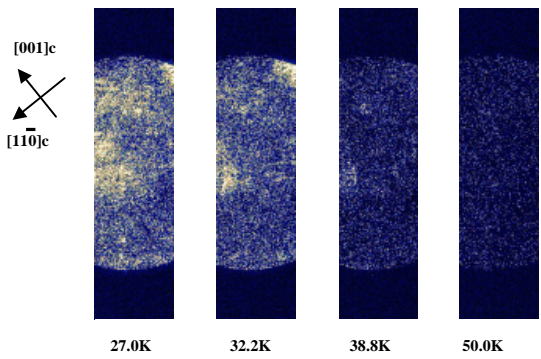

27.0K  32.2K  38.8K  50.0K

Fig.1 : SHG microscope images of STO18 in ZFC process. Bright regions indicate SHG active ones. The fundamental beam has Gaussian profile, which is a little deviated from the center of the image.

The origin of macroscopic intensity distribution could be attributed to some interference effects of SH waves, imperfections of the sample. It could be also related to macroscopic strain distribution around polar micro-regions: Recent investigation of $LiNbO_3$ discloses such stress field spreading over several 10μm near domain boundaries[7]. On the other hand, an appreciable gradient of $^{18}O$ in the sample is negligibly small, as the reaction of replacement of $^{16}O$ by $^{18}O$ is in equilibrium. Regarding the homogeneity of $^{18}O$ in the sample, we have checked it by EPMA and SIMS at room temperature. According to these results, there is no inhomogeneity in the wide plane of 7x2 $mm^2$ or the $^{16}O/^{18}O$ gradient along the thickness, with the precision 1000 A. Only a quite homogenized distribution of $^{18}O$ was observed in these experiments. In any case, the SHG microscopic observation enables us to choose the most appropriate regions in viewing a specimen and to avoid regions with defects or imperfections which produces parasitic SHG. This is important to measure precisely the integrated SH intensity from a sample.

SH intensities were measured in ZFC and field-heating (FH) processes with E=0.2kV/cm from 120K down to ~25K, and the results are shown in Fig.2.

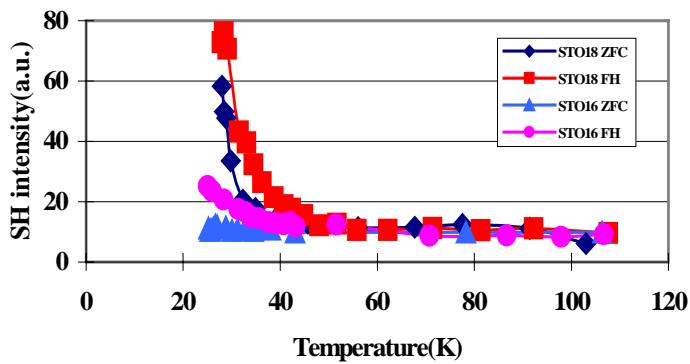

Fig.2 : Temperature dependences of SHG intensities of STO16 and STO18 on ZFC and FH after ZFC.

Here SH intensities were integrated within a selected region of SH images. In STO16, SH intensity on the ZFC is quite weak and almost temperature independent, but marked increase was observed under electric field. SH intensity on the FH process decreases gradually with the increase of temperature and vanishes at about 105K which is the critical temperature of the well-known ferroelastic phase transition.

On the other hand, SH intensity of the ZFC of STO18 is much larger than STO16, and increases steeply below about 36K. The application of electric field augments SH intensities as in STO16.

Field-induced SHGs of STO16 and STO18 are shown in Fig.3(a) and (b), respectively. In both cases, the SH intensity increases with electric field without saturation and returns to the initial zero-field value with small hysteresis when the field is switched off.

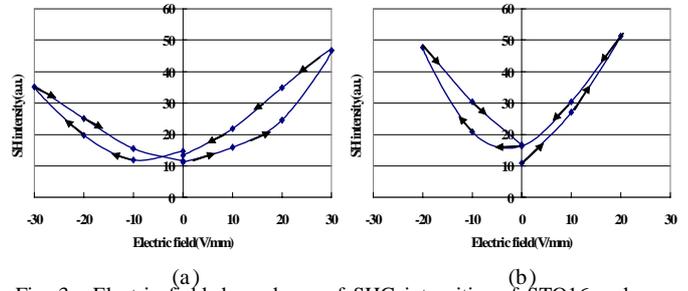

(a) (b)
Fig. 3 : Electric field dependence of SHG intensities of STO16 and STO18. (a) indicates that of STO16 at 25K, (b) STO18 at 27K.

Here it should be pointed out that the distribution of SHG active regions are still inhomogeneous under the electric field and only the total SH intensities increase in both samples. This suggests that the effect of the dipole alignment of SH active regions which are still discrete under the electric field is essential. During the SHG observations, no distinct appearance of ferroelectric domains associated with the ferroelectric phase transition was observed. These observations suggest that a ferroelectric long-range order does not prevail on both compounds and restricted to local regions, though a quantitative difference exits between STO18 and STO16.

In order to confirm the results of SHG observation of the polar order, we performed full neutron Rietveld analyses at 300 K, 50 K and 1.5 K using STO18 powders ($^{18}O$ exchange ratio=88%). The neutron experiments were performed at Laboratoire Leon Brillouin using the Orphee reactor facilities (Saclay, France). Powder diffraction patterns were collected on the high resolution two-axis goniometer 3T2 (λ=1.226 Å), using 2θ steps of 0.05° between 6° and 120°. X-ray diffraction studies in crystals and ceramics were performed using a two axis goniometer with Bragg-Brentano geometry with respectively CuKα$_1$ (InP monochromator) and CuKβ monochromatic radiation (P.G. monochromator) from a 18kW Rigaku rotating anode ; patterns were scanned through 2θ steps of 0.006° with typical counting time of 2 s up to 120 s. In each case for the low temperature experiments a He cryostat with thermal stability of 0.1 K and precision within 1 K was used. Structural refinements were carried out with the XND program[9].

At room temperature we observed a cubic Pm-3m structure, with associated normal thermal parameters. At 1.5 K and 50 K, as we have previously reported in STO16 [10,11], the superstructure peaks (Fig.4) associated with the rotation of oxygen octahedral are observed without additional superstructure peaks indicating other type of rotation along different axis.

These patterns could be very satisfactorily refined using as a starting model the tetragonal ferroelastic structure (I4/mcm space group) of STO16. The refinements converge rapidly to good values of agreements factors (at 1.5K : $R_{wp}$=4.56, $R_B$=2.22 G.O.F=1.62). The intensities of the superstructure peaks allow to measure the angle of rotation of oxygen octahedra which occurs at the ferroelastic phase transformation; at 1.5K we get a 1.97° value which is (with the experimental errors) equal to the value of 2.01 which we observed in STO16 in a previous study [11]. No anomaly in the thermal parameters which could be an indication of some possible disorder (as observed for instance in lead based relaxor compounds [12]) is evidenced.

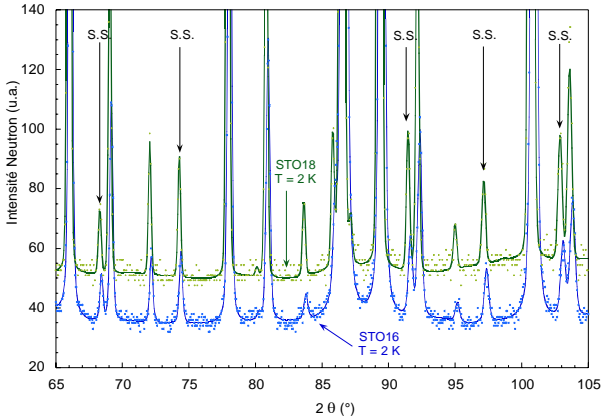

Fig.4 : Neutron diffraction profiles of STO16 and STO18 obtained at 2K. The solid lines indicate the profiles calculated by the Rietveld analysis. S.S. indicates superstructure reflections originating from the Oxygen-octahedron rotation.

High resolution X-ray experiments of powdered STO18 down to 8K reveal no additional splitting or widening from those of the tetragonal ferroelastic phase which could have indicated supplementary distortions (e.g. $BaTiO_3$-like); the value of this distortion is the same as in STO16. Nevertheless we have also tried to refine the neutron and X-ray patterns of STO18 using different ferroelectric structural models, for instance with a tetragonal but non centro-symmetric group which should allow both ferroelectric shifts along the c-tetragonal direction and the STO16-type rotations of octahedra, i.e. I4cm. However no improvement could be achieved in the agreement factors and the shifts are, within the experimental bars of 0.03A, equal to zero. The same situation was observed when we tried other models, such the monoclinic models recently observed in giant piezoelectric relaxors in the morphotropic compositions [13-16]. We also performed high resolution X-ray diffraction in single crystals of STO16 for different types of plates orientations (100, 110 and 111) with the electric field along the same direction; for these experiments we performed either ZFC down to 35K and switching on the field afterward to values up to 5kV/cm, or with FC after FH procedures. Consistently with the onset of the ferroelastic tetragonal phase, the associated Bragg peaks split at low temperatures due to the tetragonal distortion (e.g. Fig 5), but as in the STO18 experiment no additional splitting or widening is observed.

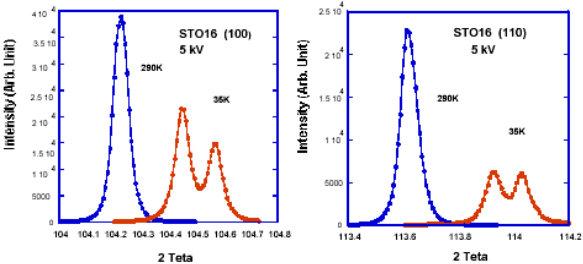

Fig.5 : Ferroelastic splittings of Xray diffraction profiles of STO16 under the electric field of 5kV/cm at 290K and 35K. (a) indicates (100) reflection, and (b) (110) reflection.

In addition we also performed in a ceramic of STO16 a FC experiment down to 10K at a high field value of 20kV/cm with the same results.

Comparison of both diffraction and SHG data is interesting as it allows to get information on the breaking of inversion symmetry at different scales. This range of sensibility slightly changes with the physical system which is probed and with the experimental conditions; in fact, with the different types of experiments of X-ray/neutron diffraction and SHG used in this study, it can be said that the former experiments probes mainly the medium and long range symmetry (roughly from some hundredth of Å up to a micrometric scale) whereas the latter is also sensible to the local order (roughly on a nanometric scale) as the SHG signal is a measure of the squared ferroelectric order parameter $<P^2>$ and its fluctuation $<\delta P^2>$. In the diffraction experiments described above we have observed in the different samples that a low temperature long range ferroelastic state which is non ferroelectric was maintained down to the lowest temperature. Clearly the combination of both high resolution X-ray and neutron diffraction studies, which has been recently successful to evidence very weak monoclinic distortions in morphotropic relaxors (see e.g. ref 16) has excluded the existence of supplementary long-ranged atomic displacements than those of the ferroelastic phase. On the other hand, SHG images have revealed at the lowest temperatures a complex and inhomogenous texture which results from the progressive appearance of polar regions in a percolative-like mechanism. However, no evidence of local polarization was observed by the diffraction experiments; in the case of lead-based perovskites the existence of short-ranged polar ordering on few hundredth of Å manifest itself in the diffraction patterns by strong temperature-dependent diffuse scatterings as well as by anomalous thermal parameters [5,12]. Such lack of observation in $SrTiO_3$ arises from two possible reasons : weaker atomic scattering factor of Sr, Ti and O, compared with that of Pb which is responsible for the diffraction signature of short range in relaxors compounds, and/or shorter length of coherence (i.e. below some hundredth of Å ) for the observed local polarisation of STO. In a similar manner , in $SrBaTiO3$ systems Lemanov et al evidenced a dipole glass behaviour of the dielectric permittivity [17] which manifests no direct signature in the diffraction patterns [10]. In STO18 the inhomogeneous texture appears to be a stable state whereas in STO16 this state is stabilized or destabilized by the application of an external electric field. The combination of both results in diffraction and SHG studies indicates that ferroelectricity in the sense of crystallographic noncentrosymmetric phase can be definitively excluded in STO16 under electric field and in STO18. However local polar ordering appears but does not develop into a true ferroelectric phase: The polar state is therefore short ranged. Similar phenomenon has been already observed in $KTaO_3$ using Raman scattering [18]

Financial supports of Grant-In-Aid for Science Research from MEXT, Grant for Development of New Technology from Shigaku-Shinkozaidan, and a Grant-in-Aid for the 21[th] COE Program (Physics of Self-organization Systems) are gratefully acknowledged.